\documentstyle[12pt]{article}
\def\lb       {\left( }
\def\rb       {\right) }
\def\lmb      {\left\{ }
\def\rmb      {\right\} }
\def\lbb     {\left[ }
\def\rbb      {\right] }
\def\comma      { \, , }
\def\period     { \, . }
\def\bra#1      { \langle \, #1 \, \vert \, }
\def\ket#1      { \, \vert \, #1 \, \rangle \, }
\def\semiket#1  { \, #1 \, \rangle \, }
\def\del        {  \partial  }
\def\half       {  {1\over 2}  }
\def\Tr      {  \mbox{Tr}  }
\def\abs#1      {  \, \vert #1 \vert \,   }
\def\bfR     { {\bf R}}
\def\bfZ     { {\bf Z}}

\def\vecii#1#2      {  \left(\begin{array}{c}#1\\#2\end{array}\right)  }
\def\veciii#1#2#3   {  \left(\begin{array}{c}#1\\#2\\#3\end{array}\right)  }
\def\matrixii#1#2#3#4            {  \left(\begin{array}{cc}#1&#2\\#3&#4
                                       \end{array}\right) }
\def\matrixiii#1#2#3#4#5#6#7#8#9 {  \left(\begin{array}{ccc}#1&#2&#3\\
                                     #4&#5&#6\\#7&#8&#9\end{array}\right)  }
\def\eqabegin         {  \begin{eqnarray}  }
\def\eqaend           {  \end{eqnarray}  }
\def\nn               {  \nonumber  }

\def\parsmallskip      {  \par\smallskip  }

\def\sectionnumbering { \setcounter{equation}{0}
         \renewcommand{\theequation}{\arabic{section}.\arabic{equation}}}
\def\mysection#1{\addtocounter{section}{1} \setcounter{subsection}{0}
                 \sectionnumbering 
    {\large\bf \par \bigskip \parsmallskip \noindent \arabic{section} \quad  
     #1  }   \par \bigskip \noindent}
\def\mysubsection#1{\addtocounter{subsection}{1}
      \par \bigskip \noindent  {\normalsize\bf
      \arabic{section}.\arabic{subsection} \quad #1  } 
   \par \medskip \noindent }
\def\csectionast#1    { \begin{center} \par \bigskip \parsmallskip 
  \noindent  {\large\bf #1  }   \par \bigskip \noindent \end{center} }
\def\alp {\alpha'}
\def\delbar {\bar{\partial}}
\def\zbar   {\bar{z}}
\def\kket#1   { \vert #1 \rangle \! \rangle }
\def\bbra#1   { \langle \! \! \langle  #1 \vert }
\def\JBH         { J_{ {\scriptscriptstyle BH  } } }
\def\MBH         { M_{ {\scriptscriptstyle BH  } } }
\def\nw   { n_{ {\scriptscriptstyle W}} }
\def\mj   { m_J }
\def\hatrp { \hat{r}_+ }
\def\hatrm { \hat{r}_- }
\begin{document}
%
\def\papertitlepage{\baselineskip 3.5ex \thispagestyle{empty}}
\def\preprinumber#1#2#3{\hfill \begin{minipage}{4.2cm}  #1
              \par\noindent #2
              \par\noindent #3
             \end{minipage}}
\renewcommand{\thefootnote}{\fnsymbol{footnote}}
%
%
\papertitlepage
\preprinumber{May 1997}{PUPT-1706}{hep-th/9705208}
\baselineskip 1.0cm 
\vspace{1.7cm}
\begin{center}
{\large\bf
  Ghost-free and Modular Invariant Spectra \\ 
    \vskip 1.5ex
    of a String in \\
$ SL(2,R) $ and Three Dimensional Black Hole Geometry
  }
\end{center}
\vskip 4ex
\begin{center}
     {\sc Yuji ~Satoh
   \footnote[2]{e-mail address: ysatoh@viper.princeton.edu} }\\
 \vskip -1ex
    {\sl Department of Physics, Princeton University \\
 \vskip -1ex
   Princeton, NJ 08544, USA}
\end{center}
\baselineskip=0.6cm
\vskip 7ex
\begin{center} {\large\bf Abstract} \end{center} 
\par \bigskip
Spectra of a string in $ SL(2,R) $ and 
three dimensional (BTZ) black hole geometry are discussed.
We consider a free field realization of $ \widehat{sl} (2,R) $ different 
from the standard ones in treatment of zero-modes. Applying this to the string 
model in $ SL(2,R) $, we show that the spectrum 
is ghost-free. The essence of the 
argument is the same as Bars' resolution to the ghost problem, 
but there are differences; 
for example, the currents do not contain logarithmic cuts. 
Moreover, we obtain a modular 
invariant partition function. This realization is also applicable 
to the analysis of 
the string in the three dimensional black hole geometry, the model of which is 
described by an orbifold of the $ SL(2,R) $ WZW model. We obtain 
ghost-free and modular invariant spectra 
for the black hole theory as well. These spectra provide examples of 
few sensible spectra of a string in non-trivial backgrounds
with curved time and, in particular,  
in a black hole background with an infinite number of propagating modes. 
\newpage
\renewcommand{\thefootnote}{\arabic{footnote}}
\setcounter{footnote}{0}
\setcounter{section}{0}
\baselineskip = 0.6 cm

\pagestyle{plain}
\setcounter{page}{1}
\mysection{Introduction}
A string in $ SL(2,R) $ 
provides one of the simplest models 
of strings in non-trivial backgrounds with curved {\it space-time}.
This model is described by the $ SL(2,R) $ WZW model
and $ SL(2,R) $ becomes an exact string background. 
In contrast to well-studied curved space cases, 
we have only a few consistent string models with curved time \cite{KKL,RT}. 
Therefore the investigation of the string in $ SL(2,R) $ 
\cite{BOFW}-\cite{Huitu}
is important as
a step to understand strings in non-trivial backgrounds.
 
The three dimensional (BTZ; Ba\~{n}ados-Teitelboim-Zanelli) 
black hole geometry \cite{BTZ} is another exact string
background with curved time. The corresponding string model is described 
by an orbifold of the $ SL(2,R) $ WZW model 
\cite{HW,Kaloper}. This model is interesting also as a model 
of quantum black holes in string theory which has an infinite number of 
propagating modes.

However, the spectrum of the string in $ SL (2,R) $ is known to contain ghosts
\cite{BOFW}-\cite{Bars}. In the orbifold model,
the original ghosts in $ SL(2,R) $ disappear,
 but a different type of ghosts appears from the twisted sectors \cite{NS}.
These analyses about ghosts are based on the standard argument about 
current algebras.

The appearance of the ghosts means that these string models are not sensible 
as they stand. However, there are arguments which support the existence of a
sensible string model in the $ SL(2,R) $ (and hence possibly 
in the BTZ black hole) 
background. First, let us consider the effective action for the S-matrix of 
the bosonic string theory. In addition to flat space, this has an 
extremal point at nearly flat three dimensional anti-de Sitter space
($ AdS_3 $), or equivalently $ SL(2,R) $, and 
is shown to be unitary at one-loop order \cite{FT}. 
Second, a $ D = 5 $ black hole solution 
in type IIB theory has the $ SL(2,R) \times SU(2) $ WZW model structure near 
the horizon \cite{Tseytlin}. So, there is a possibility of ghosts. 
Nevertheless, this black hole is mapped to a bound state of D-string and 
D5-brane. Since the resultant model is regarded as unitary,
we expect that the original model is also unitary.\footnote{
These two arguments are due to 
A.A. Tseytlin. The author thanks him for comments on these issues.} 
Third, Bars has proposed a ghost-free 
spectrum of the $ SL(2,R) $ model \cite{Bars}. By making use of 
 `modified' currents, namely, a non-standard free field realization 
of the $ \widehat{sl} (2,R) $ 
currents, he showed that the ghosts in the standard argument disappear.
Analyses of the classical string propagation indicate no pathologies 
either \cite{BOFW,Bars}.  

The standard argument about current algebras 
is well established for compact group manifolds 
but it leads to the ghosts 
for non-compact $ SL(2,R) $. Then one might think that it cannot 
be applied to non-compact cases. 
In addition, although we expect to get the three dimensional flat string 
theory in the flat limit of the $ SL(2,R) $ theory, 
this is impossible in the standard argument.
To see this, 
let us first remember that $ SL(2,R) (AdS_3) $ has a constant scalar 
curvature, which we denote by $ -6 l^{-2} $. 
In the context of string theory, this is given by $ l^2 = (k-2) \alp $
\cite{HW,NS} 
where $ (2\pi \alp)^{-1} $ is the string tension, $ k $  is the 
level of the WZW model and $ -2 $ is the second 
Casimir of $ sl(2,R) $. Thus, the flat limit is achieved by $ l $ or 
$ k \to \infty $.  Next, 
recall that the $ sl(2,R) $ algebra is given by  
\eqabegin
 \left[ J^a_0 \comma J^b_0 \right]
    &=&  i  \epsilon^{abc} \eta_{cd} J^{d}_{0}  \comma
\eqaend
where $a$-$d = 0,1,2 $ and $ \eta_{ab} = $diag $(-1,1,1)$. $ J^{1,2}_0 $ 
correspond to the non-compact direction of $ SL (2,R) $ and 
$ J^0_0 $ to the 
compact direction. The primary states 
of the standard $ SL(2,R) $ current algebra are 
represented by the states of 
the global $ sl(2,R)$ representations $ \ket{j;J} $.
Here, $-j(j+1)$ is the Casimir of $ sl(2,R) $; $ J $ is the eigenvalue of 
$ J^0_0 $, $J^2_0 $ or $ J^+_0 \equiv J^0_0 + J^1_0 $. According to the 
choice of $ J $, the representations are called elliptic, hyperbolic or 
parabolic respectively. 
Note that the primary states are labeled by two parameters 
(zero-modes). On the other hand, the base (primary) states of the flat 
three dimensional theory 
are labeled by three zero-modes (momenta), for instance, 
as $ \ket{ p^0,p^1,p^2 } $. 
Therefore, the standard $ SL(2,R)$ model cannot get to the flat theory. 
Also, it is impossible to observe how the ghosts disappear in the 
flat limit.

What is wrong in applying the standard argument to the $ SL(2,R) $ model ?
The above argument about the flat limit indicates the importance of the 
careful treatment of zero-modes. In fact, this is one of the points of 
 Bars' argument.
The importance of the zero-modes in a non-trivial background 
is also pointed out in \cite{RT}. 

Along this line of thought,  we first consider 
a free field realization of $ \widehat{sl} (2,R) $ in the following.
 This is different from the standard ones in treatment of zero-modes. 
Applying this realization to the string theory in $ SL(2,R)$, 
we show that the spectrum is ghost-free. 
The essence of the argument is the same as Bars'. However, 
our realization is different: 
Bars' modified currents are in the parabolic basis and contain 
logarithmic cuts but ours are in the hyperbolic basis and contain no 
logarithmic cuts.
The treatment of zero-modes is also different.
Moreover, we obtain a modular invariant partition function.
Our realization turns out to be useful for the analysis of the 
string in the three dimensional (and the two dimensional 
$ SL(2,R)/U(1)$ \cite{Witten}) black hole geometry. 
For the three dimensional black 
hole theory, we 
obtain ghost-free and modular invariant spectra as well.
We see that these spectra are consistent with some self-dual T-duality 
and closed time-like curves \cite{BTZ,HW} are removed 
in constructing the spectrum invariant under this T-dual 
transformation. 
This work resolves 
the ghost problem of the string theories in the $ SL (2,R) $ and the BTZ black 
hole geometry and provides examples of 
few sensible spectra of a string in non-trivial backgrounds
with curved time. In particular, this is the first time that such a 
spectrum is obtained for a string in a black hole background with an
infinite number of propagating modes.
Summary and 
brief discussion about future problems are given in the final section. 
\mysection{Spectrum of a string in $ SL(2,R) $}
\vspace*{-5ex}
\mysubsection{Free field realizations of $ \widehat{sl} (2,R) $}
We begin our discussion with the Wakimoto construction
of $ \widehat{sl} (2,R) $
\cite{Wakimoto}. It is given by a free boson $ \phi $ and the bosonic 
$ \beta $-$ \gamma $ ghosts:
\eqabegin
 i J^+(z) &=& \beta (z) \comma \nn \\
 i J^-(z) &=& \gamma^2 \beta (z) + \sqrt{2k'} \ \gamma \del \phi (z) 
      + k \del \gamma (z) \comma \label{wkmt} \\
 i J^2 (z) &=& \gamma \beta (z) + \sqrt{k'/2} \  \del \phi (z) 
 \comma \nn
\eqaend
where $ J^\pm = J^0 \pm J^1 $, $ k' \equiv k - 2 $ and
\eqabegin
  \beta (z) \gamma (w) & = & - \gamma (z) \beta (w) \ \sim \  \frac{1}{z-w}
     \comma \nn \\
     \quad \phi (z) \phi (w) & \sim &  - \ln ( z - w )
   \period 
\eqaend
It is easy to check the algebra
\eqabegin
  J^+(z) J^-(w) & \sim & \frac{-k}{(z-w)^2} + \frac{ - 2i J^2(w)}{z-w} 
     \comma \nn \\
  J^2 (z) J^\pm (w) &\sim & \frac{\pm i J^\pm (w)}{z-w }\comma \label{KM} \\
  J^2(z) J^2(w) & \sim &  \frac{k/2}{(z-w)^2} 
  \period \nn
\eqaend
In terms of the modes, this is written as 
\eqabegin
\left[ J^a_n \comma J^b_m \right]
    &=&  i  \epsilon^{ab}_{\ \ c} J^{c}_{n+m} + \frac{k}{2} n
       \eta^{ab} \delta_{m + n} \period
\eqaend
In addition, the energy-momentum tensor is given by 
\eqabegin
   T (z) &=& \frac{1}{k-2} \eta_{ab}: J^a J^b :  \nn \\
       &=&  - \half \lb \del \phi \rb^2 
    - \frac{1}{\sqrt{2k'}} \ \del^2 \phi + \beta \del \gamma
   \comma
\eqaend
and the central charge is
\eqabegin
   c &=& \lb 1 + 12 \lb 1/\sqrt{2k'} \rb ^2 \rb  +  2 
   \ = \  \frac{3k}{k-2} \period
\eqaend
At the critical value $ c = 26 $, we have $ k = 52/23 $. 
Notice that the level of the corresponding WZW model $ k $  need not 
be an integer since $ \pi_3 \lb SL(2,R) \rb = 0 $.

In the above construction, a generator to the 
non-compact direction $ J^2_0 $ is  
the Cartan subalgebra of $ SL(2,R) $ and $ J^\pm_0 $ are the step 
operators. 
Another choice of the basis is $ I^0 \equiv J^0 $ and $ I^\pm \equiv J^1 
\pm i J^2 $ in which $ I^0_0 $ is the Cartan subalgebra and $ I^\pm_0 $
are the step operators. 
 The algebra in the latter basis is 
obtained from (\ref{KM})  formally 
by $ J^\pm \to - i I^\pm $ and $ J^2 \to i I^0 $ \cite{KMS}.
The similarity to $ \widehat{su} (2) $ becomes clearer in the latter.

The base states on which the operators act are labeled by two 
zero-modes; the Casimir and, e.g., the eigenvalue of $ J^2_0$. 
Because of the deficiency of the zero-modes, 
the string model  
based on this standard realization cannot have the flat limit 
as discussed in the introduction.  
It may be  natural on physical grounds for a sensible $ SL(2,R) $ model
to have such a limit. Thus, we will try to make the $ \widehat{sl} (2,R) $ 
realization which has the flat limit by incorporating an additional zero-mode. 

For this purpose, we first bozonize the $ \beta$-$ \gamma $ ghosts. 
In principle, various bozonizations are possible because the current algebra 
is maintained as long as the OPE (operator product expansion) 
of the $ \beta$-$ \gamma $ ghosts are 
preserved.

One simple bozonization is given by two free bosons $ \phi_{0,1} $ :
\eqabegin
  \beta & = & \frac{1}{\sqrt{2}} \del \phi_+ \comma \qquad 
  \gamma \ = \ \frac{1}{\sqrt{2}}  \phi_- \comma \label{bsnz1}
\eqaend 
where $ \phi_\pm = \phi_0 \pm \phi_1 $,   
$ \phi_i(z) \phi_j (w) \sim - \eta_{ij} \ln (z-w) $ and 
$ \eta_{ij} = $ diag $ (-1,+1) $. We can readily check the current 
algebra in this realization.
In fact, the resultant currents are 
nothing but Bars' modified currents \cite{Bars}. He found these currents 
by trials and errors, but we got them in a simple way. By a careful treatment
of zero-modes, he showed that the spectrum of the $ SL(2,R) $ model using 
these currents is ghost-free. Notice that substituting (\ref{bsnz1})
into (\ref{wkmt}) yields logarithmic cuts in the currents because
$ \phi_- $ has the mode expansion 
$ \phi_- = q^{-'} -i \alpha_0^{-'} \ln z + \cdots $.
The currents are `modified' by this logarithmic term. 
We have to require that the logarithmic cuts of the currents and primary
fields have no effects in the physical sector. As a result, only particular 
combinations of the left and the right sector are allowed 
in the physical sector of the full theory.
 Also, $ J^+_0 $ is diagonal on the base states 
$ \ket{P^i} $, where $ P^i $ are the momenta 
of $ \phi_\pm $ and  $ \phi $. This means
that the representations are parabolic. 

It turns out that the representations in the hyperbolic (i.e., 
$ J^2_0 $-diagonal) basis are necessary to analyze the string theory in 
the three dimensional (and the two dimensional $ SL(2,R)/U(1) $) 
black hole geometry.
Since only certain states are allowed in Bars' realization, we cannot make the 
change from the parabolic to the hyperbolic basis. 
We are interested in the application to the black hole physics. So,  
we will consider a different realization using a different bosonization of the 
$ \beta $-$ \gamma $ ghosts in what follows. 
We will get a ghost-free and, moreover, modular 
invariant spectrum of the $ SL(2,R) $ model using this realization. 

The bosonization we then take is the standard one in \cite{FMS}.
First, we represent the $ \beta $-$ \gamma $ ghosts by 
\eqabegin
    \beta (z) = \ - e^{- \varphi_1 (z)}  \del \xi (z) \comma \quad
      \gamma (z) = \ e^{\varphi_1 (z)} \eta (z) \comma \label{BZ1}
\eqaend
where $ \varphi_1 (z) $ is a free boson, $ \eta (z) $ and $ \zeta(z) $
are a dimension-1 and a dimension-0 fermionic
field, respectively. The fermionic 
fields have the OPE
\eqabegin
  && \xi (z) \eta (w) \sim \eta (z) \xi (w)  \sim + \frac{1}{z-w}
   \period
\eqaend 
We further bosonize the fermionic fields by another free boson;
\eqabegin
  && \xi (z) =  \ e^{-\varphi_0 (z)} \comma \quad 
    \eta (z) = \ e^{\varphi_0 (z)}
   \period
\eqaend
$ \varphi_i (z) \ (i = 0,1 )$ have the OPE's
\eqabegin
  && \varphi_i (z) \varphi_j (w) \sim - \eta_{ij} \ln (z-w) \comma
\eqaend
where $ \eta_{ij}= $ diag$ (-1,1)$. Note that the zero-mode of $ \xi $ 
never appears in the $ \beta $-$ \gamma $ ghosts.

In order to get representations in the hyperbolic basis, we  
make a change of variables
\eqabegin
   X_0 &=& \sqrt{k'/k} \ \phi + \sqrt{k/2} \ \varphi_0 + k'/\sqrt{2k} \ 
   \varphi_1
   \comma  \nn \\
   X_1 &=&  \sqrt{k'/k} \ \phi - \sqrt{2/k} \ \varphi_1 
      \comma \\
  X_2 &=& \phi + \sqrt{k'/2} \lb \varphi_0 + \varphi_1 \rb
  \period \nn 
\eqaend
The new fields have the OPE's
$
   X_i(z) X_j (w) \sim - \eta_{ij} \ln (z-w)  
$
with $ \eta_{ij} = $ diag $ (-1,1,1) $. 
A similar transformation has been discussed, e.g.,
 in \cite{IK} for $ \widehat{su} (N) $. 
Consequently, the currents and the energy-momentum tensor are 
written as 
\eqabegin
   i J^\pm &=& \ e^{\mp \sqrt{2/k} X_- } 
    \del \lb \sqrt{k/2} \  X_0 \mp \sqrt{k'/2} \ X_2 \rb \comma \nn \\
   i J^2 &=& \sqrt{k/2} \ \del X_1 \comma  \label{KMcurrent}  \label{EMtensor}
    \\
   T &=& \half \del X_+ \del X_- 
     - \half \lb \del X_2 \rb^2 - \frac{1}{\sqrt{2k'}} \del^2 X_2 
   \comma \nn 
\eqaend
where $ X_\pm = X_0 \pm X_1 $.
We remark that (i) the currents have no logarithmic cuts, 
(ii) $ J^2_0 $ is diagonal on the base states 
$ \ket{p^-,p^+,p^2} $ as desired, where $ p^i $ are the momenta of 
$ X_i$ and (iii)
the energy-momentum tensor is expressed by flat light-cone free fields 
$ X_\pm $ plus a space-like free field of a Liouville type $X_2$.

Using these expressions, we find that
\eqabegin
  V^j_J  (z) & \equiv & 
   \exp \lbb i J \sqrt{2/k} \  X_- (z) + j \sqrt{2/k'} \ X_2 (z) \rbb 
   \label{primary}
\eqaend
are the primary fields of the current algebra. Actually, it follows that  
\eqabegin
  \mbox{dim. } V^j_J (z)  & = &  \frac{-j(j+1)}{k-2} \comma  \\
   J^2 (z) V^j_J (w) & \sim & \frac{J}{z-w} V^j_J (w) \comma \quad 
   J^\pm (z) V^j_J (w) \ \sim \ \frac{J \mp i j }{(z-w)} V^j_{J\pm i}(w)
  \period  \nn 
\eqaend
Also, we find the three screening operators which 
are dimension-1 Virasoro
primary fields and have no non-trivial OPE's with the currents;
\eqabegin
 \eta (z) & =  & 
   \exp \lbb  \sqrt{k/2} \ X_0 (z) - \sqrt{k'/2} \  X_2 (z) \rbb \comma \nn \\
 \tilde{\eta} (z) & = &  
   \exp \lbb - \sqrt{k/2} \ X_0 (z) - \sqrt{k'/2} \ X_2 (z) \rbb \comma  \\
  S(z) & = & 
     \del X_0(z) 
    \ e^{- \sqrt{2/k'} X_2 (z)} 
       \period \nn  
\eqaend  
These screening operators are determined up to factors and 
total derivative terms.
In particular, $ S(z) $ is expressed also as $  iJ^\pm 
  \ e^{\pm \sqrt{2/k} X_- (z) - \sqrt{2/k'} X_2 (z) } $. 
\mysubsection{Realization in an extended space}
We have made a free field realization of 
$ \widehat{sl} (2,R) $ in the basis $ J^\pm $ and $ J^2 $.
This is a 
simple application of the literature about $ \widehat{su} (2) $, e.g. 
\cite{Wakimoto,IK,NB}, and 
$ \widehat{sl} (2,R) $ in a different basis \cite{GH}. 
In the standard argument, the representation space  
is constructed as follows. First, we take the vacuum
$ \kket{ 0 } $ satisfying $ J_n^a \kket{ 0 }  = 0 $
for $ n \geq 0 $. The primary states are given by $ \kket{ j; J } \equiv 
\lim_{z \to 0} V^j_J (z) \kket{ 0 }  $. In addition, the current module 
is obtained by acting $ J_n^a $ on the primary states. The Virasoro weight is 
written as 
\eqabegin
 && L_0 = -\frac{j(j+1)}{k-2} + N  \comma 
  \label{L0KM} 
\eqaend
where $ N $ is the total grade.

Let us rephrase this in terms of the free bosons. 
First, we note that $ X_2 $ has the background charge 
$ i Q \equiv  - i \frac{1}{ \sqrt{2k'} } $. Because of $ Q $, 
the mode expansion of $ X_2 $
and the corresponding Virasoro operators are written as
\eqabegin
   X_2 (z) & = & q^2 - i(\alpha^2_0-iQ) \ln z + i \sum_{n \neq 0}
      \frac{ \alpha^2_n }{ n } z^{-n} \comma \nn \\
   L_n^{2} & = & \frac{1}{2} \sum_{m} \alpha_m^2 \alpha^2_{n-m}  
     + i Q n \alpha^2_n + \half Q^2 \delta_n
  \period
\eqaend
The total Virasoro operators are then given by
$ L_n = L_n^{\pm} + L_n^{2} $, where
\eqabegin
   && L_n^{\pm} \ = \ - \frac{1}{2} \sum_{m} \alpha_m^+ \alpha_{n-m}^-
  \comma \label{Ln}
\eqaend 
and $ \alpha_n^\pm = \alpha_n^0 \pm \alpha_n^1 $.
From the expression of $L_n$, we find that 
$ \ket{p^\pm = 0 , p^2 = i Q} $ is the $ sl_2 $-invariant vacuum.
This is nothing but the vacuum
in the standard argument $ \kket{ 0 } $. Also, from (\ref{primary}), we 
have  
\eqabegin 
&&  \kket{j;J} = \biggm{\vert} p^- = 0, \ p^+ = \sqrt{2/k} \ J , \ 
   p^2 = i(Q  - \sqrt{2/k'} \ j ) \biggm{\rangle}
    \period \label{jJ}
\eqaend
This shows the relation between 
the Casimir, $ -j(j+1) $,  and the eigenvalue of $ \alpha_0^2 $, $ p^2$;  
\eqabegin
   j & = & - \half + i \sqrt{k'/2} \ p^2 \period
\eqaend
Substituting $ p^i $ in (\ref{jJ}) into $ L_0 $  
reproduces (\ref{L0KM}). We remark that, for $ p^2 \in \bfR $,   
the above $ j $-values are precisely those of  the principal 
continuous series of the unitary $ SL (2,R) $ representations.
(For details, see, e.g., \cite{VK,NS}.)

On the entire module $ p^- $ is fixed to be zero  because
the currents shift $ p^+ $ only (see (\ref{KMcurrent})). $ J^\pm $ shift 
$ J $ by $ \pm i $. This might be curious but this is one of 
the characteristic features of the representations of $ SL(2,R) $ in 
the hyperbolic basis \cite{KMS,VK,NS}. 
First, since $ J^2_0 $ corresponds to the 
non-compact direction of $ SL(2,R) $, the spectrum of $ J $ is not discrete 
but continuous. 
So, an element (a state) of the global representation space
is given by a ``wave packet''
\eqabegin
  \ket{ \Psi } & = & \int_{-\infty}^{\infty} d J \ \Psi (J) \ket{ j;J }
  \period
\eqaend 
This is analogous to a state in a field theory using a plane wave basis 
in infinite space. The action of the operators is given by
\eqabegin
   J^2_0 \ket{ \Psi } &= & 
    \int_{-\infty}^{\infty} d J \ J \ \Psi (J) \ket{ j;J }
   \comma \nn \\
   J^+_0 \ket{ \Psi } &= & 
    \int_{-\infty}^{\infty} d J \ f_+ (J) \ \Psi (J-i) \ket{ j;J }
   \comma \\
   J^-_0 \ket{ \Psi } &= & 
    \int_{-\infty}^{\infty} d J \ f_- (J+i) \ \Psi (J+i) \ket{ j;J }
   \comma \nn
\eqaend
where $ f_\pm (J) $ play the role of the matrix elements of $ J^\pm_0 $.
The above shift of $ J $ should be understood in this way as 
that of the argument of $ \Psi (J) $.

So far, nothing is special.  Here, we will make a change from the standard
argument.  Namely, we take all $ \ket{ p^\pm, p^2  \in \bfR } $ as
base states.  This means that (i) we extend the representation space 
so that $ p^- \neq 0 $ are allowed and (ii) we concentrate on the principal 
continuous series as the global $ SL(2,R) $ representations 
and take all the representations of this type.
By this prescription, we incorporate the deficient zero-mode 
in the standard argument.\footnote{
The degrees of freedom 
of the fixed zero-mode such as $ p^- $ play roles also in different
contexts.
For example, vacua $ \ket{P^1 = i n} $, where $ P^1 $ is the momentum of 
$ \varphi_1 $ in (\ref{BZ1}) and $ n \in \bfZ $, are the 
picture-changed vacua 
with respect to the $ \beta $-$ \gamma $ system \cite{FMS}. In addition, 
the operators $ \exp ( n \sqrt{k/2} \ q^1 ) $, where $ n \in \bfZ $, 
generate the 
twists of the current algebra by integers \cite{GO}.} 
\mysubsection{Ghost-free spectrum}
Now, we are ready to show that the spectrum is ghost-free. 
First, following the arguments for the no-ghost theorem in flat 
space-time \cite{noghost}, we can prove that the module
\eqabegin
  && \prod_{l=1}^{\infty} \lb \alpha^+_{-l} \rb ^{a_l}
  \prod_{m=1}^{\infty} \lb \alpha^-_{-m} \rb ^{b_m}
  \prod_{n=1}^{\infty} \lb \alpha^2_{-n} \rb ^{c_n}
  \ket{ p^-, p^+, p^2 } 
\eqaend
contains no ghosts, where $ p^\pm, p^2 \in \bfR $ and $ a_l $, $ b_m $ and 
$ c_n $ are non-negative integers. Namely, the physical state
conditions $ ( L_n - \delta_n  ) \ket{ \psi }  = 0 $ $ (n \geq 0) $ 
remove all 
the negative-norm states in this module. 
We do not repeat the detailed argument.
Instead, it is sufficient to check that (i) the central charge of 
$ L_n $ is equal to $ 26 $, (ii) the energy-momentum tensor is hermitian, 
i.e.,  $ (L_n)^{\dag} = L_{-n} $ and 
(iii) the transverse space is positive definite 
(please remember that we have the flat light-cone ($X_\pm$) plus 
transverse ($X_2$) space). 
(i) follows from setting $ k $ to be the critical 
value $ k = 52/23 $. Once we define the hermiticity of the modes by 
$ (\alpha^i_{n})^{\dag} = \alpha^i_{-n} $, (ii) is also satisfied 
since $ Q $ is real. The oscillator 
part of $ X_2 $ causes no trouble to show (iii) because $ X_2 $ is 
space-like. But we still need to give the precise prescription of the inner
product of the zero-mode part. 
For this purpose, we first note that the conjugate of the $sl_2$-invariant 
vacuum is given by $ \bra{ p^\pm = 0, p^2 = -iQ } $. 
Then we define the conjugates of the in-states 
$ \ket{ p^2 \in \bfR } = \lim_{z \to 0} 
\ e^{ j \sqrt{2/k'} \ X_2 (z)} \ket{ i Q }  $ 
by 
$ \bra{ p^2 } = \lim_{z \to \infty} 
\bra{ - i Q } \ e^{ \bar{j} \sqrt{2/k'}  X_2 (z)}  z^{2L_0^{2}} $. 
Notice that the imaginary momentum in the $ sl_2 $-invariant vacuum 
is precisely
canceled by acting the primary fields $ V^j_J $ and that the complex 
conjugation $ j \to \bar{j} = -j-1 $  does not change the Casimir and the 
Virasoro weight. Next, we define the inner product by 
$ \bra{ p^{2'}} \semiket{ p^2 } = 2 \pi \delta (p^{2'} - p^2)  $.
This assures the positivity of the transverse space.

From the viewpoint of the original $ SL(2,R) $ model, the above statement
means that the spectrum of the $ SL(2,R) $ model is ghost-free. 
Also, by the limit $ k \to \infty $, we obtain a flat theory (although
it should be a part of the critical theory). 

It may be instructive 
to see the difference from the standard argument in detail.
First, in order to obtain the base states $ \ket{ p^\pm, p^2 \in \bfR } $ 
by acting the primary fields,
we have to prepare $ \ket{p^- \neq 0,0, iQ} $ as `vacua' 
as well as $ \kket{ 0 } = \ket{p^\pm = 0, p^2 = iQ} $. 
On these base states,  
$ J^a_0 \ket{ p^- \neq 0,0, iQ } \neq 0 $ in general.
This is similar to spontaneous symmetry breaking as Bars mentioned 
in his argument. Moreover, since $ p^- \neq 0 $, which never appeared in the 
standard argument, are allowed, the on-shell condition is changed as
\eqabegin
   && L_0  = -\half p^+ p^- - \frac{j(j+1)}{k-2} + N = 1
   \period \label{L0}
\eqaend
If $ p^- = 0 $, the $ j$-value satisfying 
this condition is real for $ N \geq 1 $ and corresponds to the discrete series
of the unitary $ SL(2,R) $ representations. 
In the standard argument, the ghosts
arise from the discrete series for $ N \gg 1 $.  
However, real $ j$-values imply purely imaginary
$ p^2 $ and such states are removed from our module. Instead, because of 
non-zero $ p^-$, states with real $ p^2 $ can be physical. 
They correspond to the principal continuous series.
We remark that the positivity of the 
transverse space becomes invalid for purely imaginary $ p^2 $. 
It is easy to find examples of negative norm states in this case from
\eqabegin
   && \lmb (p^2)^2 + 1/(2k') \rmb \bra{p^2} \semiket{p^2} 
   = \bra{p^2} 2 L_0^2 \ket{p^2} 
   = \bra{p^2} L_{1}^{2} L_{-1}^{2} \ket{p^2} \period  
\eqaend

Arguments similar to ours do not hold for compact groups
since the principal continuous series and representations in the hyperbolic
basis are characteristic of non-compact groups. We reduced our $ SL(2,R) $
theory  almost to that of three free bosons. 
It is the above characteristics that make this possible. 
Because we use the hyperbolic basis, 
the eigenvalue
of $ J^2_0 $ is continuous and has the one-to-one correspondence with the 
momentum of $ X_1 $. 
Since we take the principal continuous series at the base, 
the character of the $ SL(2,R)/SO(1,1) $ module (the coset module by  
$ J^2 $) coincides with that of two free bosons. Namely, we have 
no non-trivial null states \cite{DLP}. Thus, the sum of the characters
over all $ p^i $ for our $ SL(2,R) $ model coincides with 
that of the three free bosons with the same central charge. 
Also, for the $j$-values of the principal continuous series, 
 we find another close connection between two theories:  
on $ \ket{ p^i = 0 } $, $ V^i_J $ look like primary fields of the flat theory;
\eqabegin
  && \lim_{z \to 0} V^j_J (z) \kket{0} = \lim_{z \to 0} \ e^{i p^+ X_-}
    \ e^{i p^2 X_2} \ket{ p^i = 0 } \period
\eqaend

Moreover, we can construct the spectral generating operators like 
DDF (Del Giudice-Di Vecchia-Fubuni) operators 
in the 26-dimensional flat theory \cite{DDF,GSW}. 
The difference arises from the fact that $ \del X_2 $ is 
not a primary field  because of the background charge. To compensate this,
we use the $ b$-$c$ ghosts or the logarithmic operators $ \ln \del X_\pm $
\cite{GSW}. By making use of them, 
the spectral generating operators are written as
\eqabegin
   A^\pm_n &=& \frac{1}{\sqrt{1-4Q^2/9}} \oint \frac{d z}{2\pi i} 
   \ e^{ in X_\pm/p^\pm}
   i \lb \del X_2 - \frac{2Q}{3}  b c \rb \comma \nn \\
  \mbox{ or } \quad B^\pm_n &= & \oint \frac{d z}{2\pi i}
    \ e^{ in X_\pm/p^\pm}
   i \lb \del X_2 - Q  \ \del \ln \del X_\pm \rb \comma
\eqaend
with $ n \in \bfZ $.
In fact, since the integrands are dimension-1 primary fields, $ A^\pm_n $ and 
$ B_n^\pm $ commute with all the Virasoro operators and create physical 
states. The non-triviality of these operators is easily checked.
In addition, each set of operators has the same commutation relations
as free field oscillators;  
\eqabegin
  && \lbb A^\pm_m , A^\pm_n \rbb = \lbb B^\pm_m , B^\pm_n \rbb = m \delta_{m+n}
  \period
\eqaend 

Finally, a comment may be in order on the hermiticity of the currents and the  
spectral generating operators.
Because of the background charge, $ X_2 $ is not hermitian:
$ (X_2 (1/z) )^{\dag} = X_2(z) + 2Q \ln z $. However, we need to 
take into account
the prescription of the conjugation. Then we have 
$ \lb X_2(z) \ket{ p^2 = iQ } \rb^{\dag} = \bra{ p^2 = - iQ } X_2 (1/z) $
for the $ sl_2$-invariant vacuum. The hermiticity of the 
currents and the spectrum generating
operators should be understood in this sense.\footnote{
The contributions from the background charge in $ A^\pm_n $ and 
$ B^\pm_n $ are sprious in any case.} 

The essence of our argument is the same as Bars' \cite{Bars};
the special treatment concerning zero-modes leads to the change 
of the Virasoro condition and hence the principal continuous 
series become relevant instead of the discrete series.
Nevertheless, our currents 
contain no logarithmic cuts and the treatment of zero-modes is different.
Furthermore, using our realization, we can argue for the modular invariance 
of the spectrum and the application becomes possible to the three (and two) 
dimensional 
black holes. These are the subjects of the following sections.
\mysubsection{Modular invariance}
So far, we have focused on the left sector of the string model 
in $ SL(2,R) $. In this subsection, we introduce the right 
sector and discuss the modular invariance of the spectrum.
In what follows, $ L (R) $ implies the quantities in the left (right)
sector and tildes refer to the quantities in the right sector.

We discussed that the sum of the  
characters for the chiral sector of the $ SL(2,R) $ model 
was the same as that of the three free bosons. 
Therefore, we immediately
obtain the modular invariant partition function (1-loop vacuum 
amplitude) by identifying
the left and the right momenta as $ p^i_L = \pm p^i_R $. This means that
$ j_L = j_R $ or $ -j_R -1 $ and $ J_L = \pm J_R $ where $ j_{L(R)} $ 
and $ J_{L(R)} $ are the $ j$-values 
and the eigenvalue of $ J^2_0 (\tilde{J}^2_0)$, respectively. 
Explicitly, we find that 
\eqabegin
  Z &=& \int \frac{d^2 \tau}{\mbox{ Im } \tau}   
     \ Z_{bc} (\tau)  \ \Tr \ e^{i \tau (L_0-c_X/24) }
     \ e^{-i \bar{\tau} (\tilde{L}_0-c_X/24) } 
      \comma  \label{ptfn}
\eqaend
is modular invariant
where $ \tau $ is the modular parameter; $ \Tr $ is the trace over the 
entire module of $ X_i $; $ Z_{bc} $ is the contribution 
from the $ b$-$c$ ghosts; $ c_X = 3k/(k-2) = 26 $.
We remark that it is 
necessary to Wick-rotate $ p^0 $ as in the flat theory.\footnote{
A modular invariant of $ \widehat{sl} (2,R) $  
 has been discussed in \cite{HHRS} by including 
the degrees of freedom of zero-modes which are absent in the standard argument. The additional degrees of freedom are related to  
the twists of $ \widehat{sl} (2,R) $ by integers, the Weyl reflection of 
$ \widehat{sl} (2,R) $ or the winding around the 
compact direction of $ SL(2,R) $.}

In the above construction, the combined primary fields take the form 
\eqabegin
    V_{J;\pm J}^j (z,\zbar) & \equiv& 
    V_{J}^j(z) \tilde{V}_{\pm J}^j (\zbar) \comma
\eqaend
where $ \tilde{V}_{\pm J}^j $ is defined similarly to (\ref{primary}).
For the $j$-values, we have set $ j_L = j_R $ as usual in the WZW models. 
This is possible 
because $ j $ and $ -j-1 $ represent the same Casimir and hence 
the same representation. To take $ -j-1 = \bar{j} $  means just to take the 
conjugate.

Suppose that the combined primary fields are expressed 
by the matrix elements 
of the $ SL(2,R) $ representations which have the 
same transformation properties.\footnote{
Harmonic analysis on $ SL(2,R) $ shows that normalizable functions 
on $ SL(2,R) $ are expanded by the matrix elements of the discrete and
the principal continuous series \cite{VK}.} 
Then they are 
represented by the hypergeometric functions \cite{VK,DVV,NS}. 
The above condition $ J_L = \pm J_R $ leads to the degenerate cases,  
and the hypergeometric functions 
have singular behavior near certain points of $ SL(2,R) $.
These points correspond to the origin 
or the horizon for the two dimensional $ SL(2,R)/U(1) $ black holes
\cite{DVV}, whereas they correspond to the inner or the outer horizon
of the three dimensional black holes \cite{NS}.
\mysection{Spectrum of a string in three dimensional black hole \\ 
   \hspace*{4ex} geometry }
In this section, we turn to the application 
 to the string theory in the three dimensional black hole geometry. 
Actually, this was important part of our motivation to investigate 
the $  SL(2,R) $ theory. The application is straightforward
and we obtain ghost-free and modular invariant spectra again.
\mysubsection{Three dimensional black holes}
Let us begin with a brief review of the three dimensional black holes 
\cite{BTZ} in the context of string theory \cite{HW}-\cite{NS}.

The three dimensional black holes are solutions to the vacuum 
Einstein equations
with the cosmological constant $ -l^{-2} $.
The simplest way to obtain the black hole geometry
is to start from the three dimensional anti-de Sitter space ($AdS_3$), 
or equivalently, 
the $ SL(2,R) $ group manifold \cite{BTZ}.\footnote{
In this section, we consider the universal covering group (space) of 
$ SL(2,R) $ ($ AdS_3 $). The argument in Section 2 holds without change.} 
In a parametrization, the metric of (a part of) $ AdS_3 $ 
is written as 
\eqabegin
  ds^2 &=& - \left( \frac{r^2}{l^2} - \MBH \right) d t^2 - \JBH d t d \varphi 
   + r^2  d \varphi^2 
     + \left( \frac{r^2}{l^2} - \MBH 
    + \frac{\JBH ^2}{4r^2}\right)^{-1} d r^2 
     \comma  \nn \\
   && 
\eqaend
where $ -\infty < t, \varphi < \infty $, $ 0 \leq r < \infty $; 
$ \MBH $ and $ \JBH $ are some parameters. The black hole geometry 
is obtained  by identifying 
$ \varphi + 2 \pi $ with $ \varphi $.
In the following, we denote this identification by $ Z_\varphi $. The 
above two parameters, $\MBH$ and $\JBH$ represent the mass and 
the angular momentum of the black hole. If we express them as
 $ l^2 \MBH = r_+^2 + r_-^2 $ and $ l \JBH =  2 r_+ r_-$ for 
$ r_+  \geq r_- \geq 0 $, then $ r = r_\pm $ correspond to the location of 
the outer and the inner horizon, respectively. The variables $ t $, $ r $ and 
$ \varphi $ represent the time, the radial and the angle coordinate.

The string theory in this geometry is described by 
the $ SL(2,R)/Z _\varphi $ orbifold of the $ SL(2,R) $ WZW model
\cite{HW,Kaloper}. Thus, the black hole geometry provides 
an exact and simple string background with curved time. 
The model is interesting both 
as a string model in non-trivial backgrounds and as a model of 
quantum black holes in string theory.
However, a detailed analysis based on the standard argument about 
the current algebras showed that the model contains ghosts as mentioned 
in the introduction \cite{NS}. 
The ghosts in the black hole model are different from those in the untwisted 
model and originate from the twisted sectors.
\mysubsection{Application of the new realization}
Now we discuss the application of the new realization.
In order to analyze the spectrum of the orbifold model by $ Z _\varphi $, 
we need to construct twist (winding) operators. 
Although our realization is different 
from the standard ones, the basic strategy is the same. 
Thus, we follow the argument in \cite{GPS} and \cite{NS}. 
First, we recall that the coordinates $ \varphi $ and $ t $ are
expressed by 
analogs of the Euler angles $ \theta_L $ and $ \theta_R $ as \cite{NS}  
\eqabegin
   \varphi & = & \half \lb \frac{\theta_L}{\Delta_-} + 
  \frac{\theta_R}{\Delta_+} 
  \rb \comma  \quad 
   t/l \ = \ \half \lb \frac{\theta_L}{\Delta_-} - \frac{\theta_R}{\Delta_+} 
  \rb \comma 
\eqaend
where $ \Delta_\pm = \hatrp \pm \hatrm \equiv (r_+ \pm r_-)/l $. 
The untwisted 
primary fields $ V^{j_L}_{J_L} (z) \ (\tilde{V}^{j_R}_{J_R} (\zbar)) $ 
should have the $ \theta_L $ ($ \theta_R $) dependence as 
$ e^{-iJ_L \theta_L} \ (e^{-iJ_R \theta_R}) $ \cite{NS}.
Next, we decompose $ \theta_{L,R} $ into the free field parts 
and the non-free field parts as 
\eqabegin
  && \theta_L = \theta_{L}^F (z) + \theta_{L}^{NF} (z,\zbar)
   \comma \quad  
   \theta_R = \theta_{R}^F (\zbar) + \theta_{R}^{NF} (z,\zbar)
  \period
\eqaend
Note that $ \theta_L^F $ is holomorphic and 
$ \theta_R^F $ is anti-holomorphic.  In terms of  
these free fields, $ J^2 (z) $ 
and $ \tilde{J}^2(\zbar) $ are written as $ i J^2 = (k/2) \del \theta_L^F $, 
$ i \tilde{J}_R^F = (k/2) \del \theta_R^F $ \cite{NS}.
Comparing these expressions with $ J^2 $ in (\ref{EMtensor}) yields  
\eqabegin
    && X_1(z) = \sqrt{k/2} \ \theta_L^F (z) \comma \quad 
       \tilde{X}_1(\zbar) = \sqrt{k/2} \ \theta_R^F (\zbar) 
   \period
\eqaend
Then, we find that the twist operator representing $ n$-fold winding 
is given by
\eqabegin
   W(z,\zbar;\nw) &=& \exp \lmb i  \lb  \mu_L X_1(z) 
    + \mu_R  \tilde{X}_1 (\zbar) \rb \rmb \comma 
\eqaend
where $
      \mu_L = \nw \Delta_- \sqrt{k/2} \comma 
       \mu_R = - \nw \Delta_+ \sqrt{k/2}   \period
      $
In fact, this operator has the OPE's
\eqabegin
   && X_1(z) W(0,0;\nw) \sim - i \nw \Delta_- \sqrt{k/2} \  \ln z \ W(0,0;\nw)
   \comma \nn \\
     && \tilde{X}_1(\zbar) W(0,0;\nw) \sim i \nw \Delta_+ \sqrt{k/2} \ 
  \ln \zbar
  \ W(0,0;\nw)
  \comma  
\eqaend
and this means that $ \delta \varphi = 2 \pi \nw $ and $ \delta t = 0 $ under 
$ \sigma \to \sigma + 2 \pi $ where $ z = e^{\tau +i \sigma }$.
A general primary field is obtained by combining 
an untwisted primary field and $ W (z,\zbar;\nw) $. As a result, 
the eigenvalue of $ \alpha_0^1 $($ \tilde{\alpha}_0^1$) 
of the twisted primary field becomes 
$ p^{1'}_{L(R)} = p^1_{L(R)} + \mu_{L(R)} $ 
where $ p^1_{L(R)} = - \sqrt{2/k} \ J_{L(R)}$ is that of the untwisted part.

Next, we solve the level matching condition $ L_0 - \tilde{L}_0 \in \bfZ $.
Here we will set the momenta of the untwisted part to be $ p^i_L = \pm p^i_R $
as in the previous section because the untwisted 
model was sensible for these cases. Also, it turns out that we obtain 
the sensible orbifold model starting from this untwisted part. 
We will see that 
the twist with respect to $ Z_\varphi $ becomes similar to 
the toroidal compactification.

The difference from the analysis in \cite{NS} arises because 
(i) the Virasoro weights $ L_0 $ 
and $ \tilde{L}_0 $ take the forms as (\ref{L0}), e.g., 
$ 2 L_0 = -(p^0)^2 + (p^{1'})^2 -2j(j+1)/k' + 2N $  and (ii) 
the left and right zero-modes of the untwisted part 
are connected by $ p^i_L = \pm p^i_R $. According to
$ p^1_L = \pm p^1_R $, we have two cases.
First, we consider  $ p^1_L = p^1_R \equiv p^1 $ case. Then, 
the level matching leads to the condition 
$ 2 \hatrp \nw \lb \sqrt{k/2} \ p^1 -(k/2)  \hatrm \nw \rb = \mj $, 
where $ \mj $ is an integer. Furthermore, this 
plus the closure of the OPE give 
\eqabegin
  \sqrt{k/2} \ p^1 
   &=& \half  \lb \frac{\mj}{\hatrp} +  k \hatrm \nw \rb \comma 
\eqaend
in other words,
\eqabegin
  \sqrt{k/2} \ p^{1'}_L & = & \half \lb \frac{\mj}{\hatrp} + k \hatrp \nw \rb  
   \comma \quad 
   \sqrt{k/2} \ p^{1'}_R \ = \ \half 
      \lb \frac{\mj}{\hatrp} - k \hatrp \nw \rb  \period
  \label{p1}
\eqaend
When the above condition is satisfied, the primary field is invariant under
\eqabegin
   \delta \theta_L^F  = \delta \theta_L^{NF} = \pi \Delta_- \comma \quad 
   \delta \theta_R^F  = \delta \theta_R^{NF} = \pi \Delta_+ \period 
\eqaend
This follows from the $ \theta_{L,R} $-dependence of the untwisted part 
and $ \theta_{L,R}^F $-dependence of the twist operator, and  
ensures the single-valuedness under $ \delta \varphi = 2 \pi $.
The non-free parts $ \theta_{L,R}^{NF} $ appear in the orbifolding 
in this way as in \cite{GPS} and \cite{NS}. The states corresponding 
to the primary fields are obtained in the usual way.

The discussion for $ p^1_L = - p^1_R $ case 
is similar, and we do not repeat it.\footnote{
In the flat limit $ k $ (or $ l $) $ \to \infty $, we have 
$ i J^2 \sim (k/2) \del \lb \theta_L - \theta_R \rb $ and 
$ i \tilde{J}^2 \sim - (k/2) \delbar \lb \theta_L - \theta_R \rb $ \cite{NS}.
 So, this latter case smoothly leads to the flat limit.} 
The results are obtained simply by the exchange 
$ \hatrp \leftrightarrow \hatrm $ and appropriate changes of signs. 

Consequently, 
the effects of the twist by $ Z_\varphi $ are summarized
in discretizing the eigenvalue of $ \alpha^1_0 $ and $ \tilde{\alpha}_0^1$
as, e.g.,  in (\ref{p1}). Therefore, we can readily confirm that the 
spectrum for each case is 
ghost-free and modular invariant. First, once we restrict ourselves to 
the left or the right sector, the spectrum of each sector is just 
a subset of the spectrum of the untwisted model. Thus, we can prove 
the no-ghost theorem in each sector and hence that of the full theory.
It is easy to  see that the examples of the ghosts in \cite{NS}
disappear.
Second, $ p^{1'}_{L,R} $  are discretized in the same way 
as the momenta of the 
troidally compactified field. So, we find that the replacement
of the integral over $ p^1 \equiv p^1_L = \pm p^1_R $ in (\ref{ptfn}) 
by the summation 
$ \sum_{\nw,\mj \in \bfZ}$ does not break the modular invariance of the 
partition function. Also, in order to get the physical states, we 
can make use of the same  
spectral generating operators as in the untwisted model.

For the three dimensional black holes, there is a self-dual
T-dual transformation $({\cal T})$ \cite{NS} which is caused by 
$ \theta_R \to - \theta_R $ \cite{GRV}. 
The flip of the sign of $ p^1_R $ corresponds to $ {\cal T} $.  
This is consistent with the above result since
$ {\cal T} $ is interpreted  also as the exchanges  
$ \hatrp \leftrightarrow \hatrm $ and $ r^2 \leftrightarrow -r^2 + l^2 \MBH $.
Although there is no rigorous proof that T-dual transformations
along non-compact directions are 
the exact symmetries of CFT's, it might be natural to suppose that
that is the case. Then the partition function should be 
invariant under $ {\cal T} $. Such an invariant partition function
is obtained   
simply by summing  up the two sectors $ p^1_L = \pm p^1_R $.\footnote{
$ \MBH $ and $ \JBH $ are also $ {\cal T} $-invariant.}
The resultant spectrum is regarded as  that of the model
further twisted by this $ \bfZ_2 ({\cal T}) $ symmetry. 
Thus, in this construction the region $ r^2 < 0 $ 
where closed time-like curves exists \cite{BTZ,HW} is truncated 
because of $ r^2 \leftrightarrow -r^2 + l^2 \MBH $ 
(at the expense of the additional truncated region 
$ 0 \leq r^2/l^2 < \MBH/2$). 
This shows a way to remove the closed time-like curves in 
the three dimensional black holes. 
\mysection{Summary and discussion}
In this paper, we discussed a free field realization of 
$ \widehat{sl} (2,R) $, which was different from the standard ones 
in the treatment of zero-modes. This gave a resolution to the ghost problem
of the string theory in $ SL(2,R) $. The essence of the argument was the 
same as Bars' which had given a resolution to this problem. However, 
our currents did not have logarithmic cuts and the treatment of zero-modes
was also different. Moreover, we obtained a modular invariant 
partition function.
Our realization was useful for the 
analysis of the string in the three dimensional black hole geometry as well.
The model was described by an orbifold of the $ SL(2,R) $ WZW model.
By a simple application, we again obtained ghost-free and modular 
invariant spectra for the black hole theory. 
These spectra were consistent with some self-dual T-duality and 
we saw that closed time-like curves were removed 
in constructing the spectrum invariant under this T-dual 
transformation. 

Each spectrum obtained here provides one of few 
sensible spectra of a string in non-trivial backgrounds
with curved time (at least so far). Furthermore, this is the first time that
such a sensible spectrum is obtained for a string theory in a black hole 
background with an infinite number of propagating modes. 

The argument in this paper may have a wide variety of applications to string
models in non-trivial backgrounds containing non-compact group manifolds.
In particular, the application to the $ SL(2,R)/U(1) $ black holes is 
straightforward because, for the lorentzian black holes, the model 
is described by 
the coset of the $ SL(2,R) $ WZW model by $ J^2 (z) \pm \tilde{J}^2 (\zbar) 
\ (i\del X_1(z) \pm i\del \tilde{X}_1 (\zbar)) $.

Finally, let us discuss the remaining problems. First, we need to 
check that the spectrum is closed at loop orders, namely, when 
interactions are introduced. This is also necessary to assure the unitarity
of the model. For this purpose, we have to analyze the fusion rules. This 
requires careful treatment of screening operators as in other free field 
systems including background charges. Calculation of correlators 
is closely related. 
One might be able to derive the fusion rules from the 
modular property of the partition function \cite{Verlinde,MS}. A naive 
application of the formula in \cite{Verlinde} leads to  
$ \phi_{\vec{p}} \times \phi_{\vec{p'}} = \phi_{ \vec{p}+\vec{p'} } $,
where  
$ \phi_{\vec{p}} $ are the chiral primary fields with momenta of $ X_i $, 
$ \vec{p} - i \vec{Q} = (p^0,p^1,p^2 - iQ) $; $ p^i \in \bfR $.
Nevertheless, there is no proof of this formula for CFT's with 
an infinite number of primary fields to the author's knowledge. 
Second, it is important to consider 
how to extract physical information. What we should do first is to calculate 
correlation functions. However, this may not be enough to understand black 
hole physics such as Hawking radiation and black hole entropy. 
Recently, these issues are intensively analyzed by using D-branes. 
It is interesting to approach these problems in a different way 
using our model.
\newpage
\csectionast{Acknowledgements}
The author would like to thank I. Bars, S. Hirano, M. Natsuume and, 
especially, M. Kato for useful discussions. He would also 
like to acknowledge useful comments from A. A. Tseytlin on \cite{NS}.
This work is supported in part by JSPS Postdoctoral Fellowships for 
Research Abroad.  
%
%
\baselineskip=0.5cm
\def\thebibliography#1{\list
 {[\arabic{enumi}]}{\settowidth\labelwidth{[#1]}\leftmargin\labelwidth
  \advance\leftmargin\labelsep
  \usecounter{enumi}}
  \def\newblock{\hskip .11em plus .33em minus .07em}
  \sloppy\clubpenalty4000\widowpenalty4000
  \sfcode`\.=1000\relax}
 \let\endthebibliography=\endlist
\csectionast{References}
%

%
\end{document}